\documentclass[12pt]{iopart}
\usepackage{graphicx}
\usepackage{dcolumn}
\usepackage{bm}
\usepackage{epsf}

\begin{document}

\title{Electron mass shift in nonthermal systems}
\author{P L Hagelstein$^1$, I U Chaudhary$^2$}

\address{$^1$ Research Laboratory of Electronics, 
Massachusetts Institute of Technology, 
Cambridge, MA 02139,USA}
\ead{plh@mit.edu}

\address{$^2$ Research Laboratory of Electronics, 
Massachusetts Institute of Technology, 
Cambridge, MA 02139,USA}
\ead{irfanc@mit.edu}

\begin{abstract}

The electron mass is known to be sensitive to local fluctuations in the
electromagnetic field, and undergoes a small shift in a thermal field.
It was claimed recently that a very large electron mass shift should
be expected near the surface of a metal hydride [{\it Eur. Phys. J. C}, {\bf 46} 107 (2006)].
We examine the shift using a formulation based on the Coulomb gauge, which leads to a much smaller shift.
The maximization of the electron mass shift under nonequilibrium conditions
seems nonetheless to be an interesting problem.
We consider a scheme in which a current in a hollow wire produces a large vector
potential in the wire center.
Fluctuations in an LC circuit with nearly matched loss and gain 
can produce large current fluctuations; and these can increase the
electron mass shift by orders of magnitude over its room temperature value.
\end{abstract}


\maketitle

\section{Introduction}
\label{sec:intro}

  The problem of the electron self-energy has been of interest since the early days
of quantum field theory; most importantly in the case of the vacuum \cite{Feynman} and 
atoms \cite{Bethe,Desiderio1971,Mohr1974}.
Subsequently, there has been interest in the electron self-energy under a variety of
conditions; 
in a strong magnetic field \cite{Parle1987};
in an intense laser radiation field \cite{Eberly1966,Reiss1979};
and in a thermal radiation field \cite{Levinson1985,Barton1990}.
Such problems have provided theorists with a rich opportunity for substantive theoretical
developments \cite{Donoghue1985}. 
One of the low-order terms that results from QED is a mass shift.
The mass shift due to a thermal field under readily accessible conditions is very small, 
but an experimental observation has been reported \cite{Hollberg1984}.
In the case of an intense laser field, the mass shift can be much greater; however, under these
conditions other processes, such as multiphoton ionization, occur \cite{Reiss1990}.

Our interest in this problem generally was stimulated by a recent paper by Widom and Larsen \cite{Widom2006}.
In this paper, the authors propose that a very large mass shift can be obtained near the surface of a 
metal hydride under nonequilibrium conditions.
According to Widom and Larsen, the electron mass shift can be in the MeV range.

Of course, a mass shift this large is unexpected and unprecedented.
To develop such a large mass shift, intuition suggests that the electron must interact 
with the local environment with at least a comparable interaction strength.
Under the relatively benign environment of a metal hydride, it is difficult to understand why 
such large interactions should occur.
If there existed such strong dynamical fluctuations, one should expect multiphoton ionization
as occurs in intense laser field; but generally no such effects are usually observed.
Consequently, we are motivated to examine the model in order to better understand the
problem.

In their paper, Widom and Larsen obtain a mass shift formula in a form that is Lorentz invariant 
and gauge free.
A specific numerical example is given in which the electric field is estimated from a
simplified model which is based on the electric field due to oscillating protons at
the metal hydride surface.
Of interest was whether the Widom and Larsen result could be confirmed in a different
formalism in which a specific gauge is specified.
Results for observable quantities must be independent of the choice of
gauge in the case of a complete computation where all effects are taken into account. 
It is well known that the choice of gauge can produce different answers in
practical computations where the computation is approximate, or not complete in this sense
\cite{Reiss1979,Lindroth1989}.
Since the Coulomb gauge is widely used, we adopted it for this purpose.
We find that the mass shift estimated using this approach for their example is lower by the
fourth power of the ratio of the proton velocity to the speed of light.
In this example, the Coulomb gauge result is lower by about eighteen orders of magnitude.
Since local electrons have higher velocities, one would expect a much larger electron-electron contribution,
especially if a significant current was present.
However, any such effects are trivial in comparison with Coulomb interactions between
electrons and ions that occur in a metal hydride.

Nevertheless, an issue underlying the Widom and Larsen paper remains of interest.
Can a large mass shift be produced somehow under nonequilibrium conditions, 
without using an intense laser field, and under conditions where other processes, such as 
multiphoton ionization, are avoided?
To this end, we consider an idealized physical situation (conditions in the center of
a hollow conductor carrying a large current) in which we seek to create a very large
vector potential and induce fluctuations that would maximize the mass shift.
If the frequency components remain sufficiently low, multiphoton transitions and ionization
should be minimized.
We find that a small electron mass shift can be generated using this approach, and 
the effect should be detectable through the observation of lines shifts for transitions
involving weakly bound electrons.

\newpage
\section{Idealized model}

It is often useful to have a highly simplified model in order to gain intuition about
an effect.
In this case we can take advantage of a similar one that has been used for this purpose previously \cite{Panella1990}.
Consider an electron in free space interacting with a transverse field according to

\begin{equation}
\hat{H}
~=~
{\boldsymbol \alpha} \cdot c {\bf p} 
+
\beta mc^2
-
{e \over c} {\boldsymbol \alpha}  \cdot \hat{\bf A}
\end{equation}

\noindent
The energy then depends on the transverse radiation field 
through

\begin{equation}
E^2 
~=~
\langle \hat{H}^2 \rangle
~=~ 
(mc^2)^2
+
\langle | c{\bf p} - {e \over c} \hat{\bf A}|^2 \rangle
-
{e^2 \over c^2}
\langle |\hat{\bf A}|^2 \rangle_0
\end{equation} 

\noindent
where we subtract out the vacuum contribution to the fluctuations, since it is already
taken into account in the mass $m$.

Assuming an approximate product wavefunction in which the electron and radiation field are taken to
be independent, we obtain

\begin{equation}
E^2 
~=~ 
(mc^2)^2
+
c^2 \langle | {\bf p} |^2 \rangle
-
e \langle {\bf p} \rangle \cdot \langle \hat{\bf A} \rangle
-
e \langle \hat{\bf A} \rangle \cdot \langle {\bf p} \rangle
+
{e^2 \over c^2} [ \langle |\hat{\bf A}|^2 \rangle - \langle |\hat{\bf A}|^2 \rangle_0]
\end{equation} 

\noindent
We introduce a shifted momentum

\begin{equation}
\langle {\bf p}' \rangle
~=~
\langle {\bf p} \rangle
-
{e \over c}
\langle \hat{\bf A} \rangle
\end{equation}

\noindent
to obtain

\begin{equation}
E^2 
~=~ 
(mc^2)^2
+
c^2 \langle | {\bf p}' |^2 \rangle
+
{e^2 \over c^2} ( \langle |\hat{\bf A}|^2 \rangle - \langle |\hat{\bf A}|^2 \rangle_0 - |\langle \hat{\bf A} \rangle|^2 )
\end{equation}

\noindent
From this we can identify the dressed mass in terms of electromagnetic field fluctuations according to

\begin{equation}
(m^*)^2 
~=~ 
m^2 +  
{e^2 \over  c^4} ( \langle |\hat{\bf A}|^2 \rangle - \langle |\hat{\bf A}|^2 \rangle_0 - |\langle \hat{\bf A} \rangle|^2 )
\end{equation}

\noindent
The mass shift $\delta m$ is then

\begin{equation}
\delta m ~=~ 
{e^2 \over  2 m c^4} ( \langle |\hat{\bf A}|^2 \rangle - \langle |\hat{\bf A}|^2 \rangle_0 - |\langle \hat{\bf A} \rangle|^2 )
\label{shift}
\end{equation}

\noindent
under the assumption that $\delta m \ll m$.

In the event that the local radiation field is a blackbody, 
the expectation value of the potential vector $\langle \hat{\bf A} \rangle$ is zero,
and one obtains \cite{Levinson1985,Barton1990}

\begin{equation}
{\delta m \over m}
~=~ 
{\pi \alpha \over 3} \left [ {k_BT \over mc^2} \right ]^2
\label{thermalshift}
\end{equation}

\newpage
\section{Mass shift in terms of current sources}

If the system is not in thermal equilibrium, we require an expression for the field fluctuations in terms of sources responsible
for the local fields.
For this, we are guided by the classical problem.
The classical vector potential in the Coulomb gauge satisfies

\begin{equation}
-\nabla \times \left [ {1 \over \mu} \nabla \times {\bf A}({\bf r},\omega) \right ]
+
{\omega^2 \epsilon \over c^2}  {\bf A}({\bf r},\omega)  
~=~ 
-{4 \pi \over c} {\bf j}({\bf r},\omega)
\end{equation}

\noindent
subject to

\begin{equation}
\nabla \cdot {\bf A}({\bf r},\omega) ~=~ 0
\end{equation}

\noindent
which defines the Coulomb gauge.
This subsidiary condition can omitted if we replace the current density by ${\bf j}_T$,
where ${\bf j}_T$ is the transverse part of the current density \cite{Jackson}.
The classical vector potential arising from sources can be constructed from those sources according to

\begin{equation}
{\bf A}({\bf r},\omega) 
~=~ 
{1 \over c} \int d^3 {\bf r}'~ G({\bf r},{\bf r}';\omega)~ {\bf j}_T({\bf r}',\omega)
\end{equation}

\noindent
where the Green's function $G({\bf r},{\bf r}',\omega)$ satisfies

\begin{equation}
-\nabla \times \left [ {1 \over \mu} \nabla \times G({\bf r},{\bf r}';\omega) \right ]
+
{\omega^2 \epsilon \over c^2} G({\bf r},{\bf r}';\omega)  
~=~ 
- 4 \pi \delta^3({\bf r}- {\bf r}')
\end{equation}

\noindent
For simplicity we assume spatial uniformity, in which case the Green's function is
a scalar.
The analogous Heisenberg operators satisfy similar relations, which allows us to write

\begin{equation}
\hat{\bf A}({\bf r},\omega) 
~=~ 
{1 \over c} \int d^3 {\bf r}' ~G({\bf r},{\bf r}';\omega)~  \hat{\bf j}_T({\bf r}',\omega)
\end{equation}

\noindent
We can then relate the electromagnetic field fluctuations to fluctuations in the source according to

{\small

$$
\langle \hat{\bf A}({\bf r,\omega}) \cdot \hat{\bf A}({\bf r,\omega}) \rangle
- 
\langle \hat{\bf A}({\bf r,\omega}) \rangle \cdot \langle \hat{\bf A}({\bf r,\omega}) \rangle
~=~
{1 \over c^2}
\int d^3 {\bf r}'
\int d^3 {\bf r}''~
G({\bf r},{\bf r}' ;\omega)
G({\bf r},{\bf r}'';\omega)
$$
\begin{equation}
\ \ \ \ \ \ \ \ \ \ \ \ \ \ \ \ \ \ \ \ \ \ \ \ \
\bigg [
\langle 
\hat{\bf j}_T({\bf r}',\omega)
\cdot
\hat{\bf j}_T({\bf r}'',\omega)
\rangle
-
\langle 
\hat{\bf j}_T({\bf r}',\omega)
\rangle
\cdot
\langle
\hat{\bf j}_T({\bf r}'',\omega)
\rangle
\bigg ]
\end{equation}

}

\noindent
A related approach was used in \cite{Greffet2005} for electromagnetic field fluctuations near surfaces.
Hence, we may write for the relative mass shift

{\small

$$
{\delta m \over m}
~=~
\int d \omega~
{1 \over 2}
\left ( {e \over mc^3} \right )^2
\int d^3 {\bf r}'
\int d^3 {\bf r}''~
G({\bf r},{\bf r}';\omega)
G({\bf r},{\bf r}'',\omega)
\ \ \ \ \ \ \ \ \ \ \ \ \ \ \ \ \ \ \ \ \ \ \ \ \ \ \ \ \ 
$$
\begin{equation}
\ \ \ \ \ \ \ \ \ \ \ \ \ \ \ \ \ \ \ \ \ \ \ \ \
\bigg [
\langle 
\hat{\bf j}_T({\bf r}',\omega)
\cdot
\hat{\bf j}_T({\bf r}'',\omega)
\rangle
-
\langle 
\hat{\bf j}_T({\bf r}',\omega)
\rangle
\cdot
\langle
\hat{\bf j}_T({\bf r}'',\omega)
\rangle
\bigg ]
\end{equation}

}

\newpage

\section{Mass shift in a metal hydride}

As discussed in the Introduction, 
Widom and Larsen have identified metal hydrides as an environment in which the mass shift
can become large \cite{Widom2006}.
Electromagnetic field fluctuations in the vicinity of a metal surface have been studied
previously \cite{Henkel2005}, and significant near-surface enhancements are reported \cite{Henkel2000}.
However, the mass shift estimate reported in \cite{Widom2006} seems to be larger than
what we would expect, so in this section we examine the model used.

\subsection{Widom-Larsen model}

To obtain an estimate for the mass shift, these authors have expressed the dressed
mass (translated into our notation) as

\begin{equation}
{m^* \over m} 
~=~
\sqrt{
1 
+ 
\left ( {e \over m c^2} \right )^2 
\overline{A^\mu A_\mu} 
}
\end{equation}

\noindent
which is developed into 

\begin{equation}
{m^* \over m} 
~=~
\sqrt{
1 
+ 
{|\overline{\bf E}|^2 \over {\cal{E}}^2}}
\label{gaugeinv}
\end{equation}

\noindent
with

\begin{equation}
{\cal{E}} ~=~ \left | {mc \tilde{\Omega} \over e} \right |
\end{equation}

\noindent
with $\tilde{\Omega}$ the local plasma frequency.  According to Widom and Larsen, their Equation (16) 
[Equation (\ref{gaugeinv}) here]
is ``an obviously gauge invariant result.''

To develop a quantitative estimate for the magnitude of the electric field fluctuations,
Widom and Larsen consider oscillations of a proton in a local pocket of electronic charge
density $-|e| \tilde{n}$.
Using Gauss's law, they obtain an estimate for the electric field fluctuations

\begin{equation}
\sqrt{|\overline{\bf E}|^2 }
~\approx~
{4 e \sqrt{|{\bf u}|^2} \over 3 a_0^3}
\end{equation}

\noindent
where ${\bf u}$ is the displacement of the proton monolayer and $a_0$ is the Bohr radius.
The estimates that result from this approach lead to estimates for the dressed mass that
can be enormous.
According to their Equation (20), they find

\begin{equation}
{m^* \over m} 
~\approx~
20.6
\end{equation}

\noindent
Such a large estimate for the mass shift provided us with the motivation to examine the model.

\subsection{Electric field operators}

  To make progress, we would like to think about the mass shift in terms of the electric field operator.
We begin by considering the classical electric field, which can be separated into longitudinal and
transverse pieces

\begin{equation}
{\bf E} ~=~ {\bf E}_L + {\bf E}_T
\end{equation}

\noindent
which satisfy

\begin{equation}
\nabla \times {\bf E}_L ~=~ 0
\ \ \ \ \ \ \ \ \ \
\nabla \cdot {\bf E}_T ~=~ 0
\end{equation}

\noindent
The transverse part is related to the vector potential through

\begin{equation}
{\bf E}_T ~=~ - {1 \over c} {\partial {\bf A} \over \partial t}
\end{equation}

\noindent
The analogous Heisenberg operators satisfy the same relation, so we may write

\begin{equation}
 \hat{\bf E}_T({\bf r},t) 
~=~ 
- {1 \over c} {\partial  \hat{\bf A}({\bf r},t) \over \partial t}
\end{equation}

\noindent
We can recast the mass shift formula in terms of the transverse electric field
operator by using the Fourier transform version of this relation.

\begin{equation}
\delta m ~=~ 
{e^2 \over  2 m c^2} 
\int {d \omega \over \omega^2}
\bigg [ 
\langle |\hat{\bf E}_T(\omega)|^2 \rangle - \langle |\hat{\bf E}_T(\omega)|^2 \rangle_0 - |\langle \hat{\bf E}_T(\omega) \rangle|^2 
\bigg ]
\end{equation}

In the formulation of Widom and Larsen \cite{Widom2006}, the appearance of 
the full electric field operator in their mass shift formula is what makes
their gauge invariant formulation different from the Coulomb gauge approach
under discussion here.
Since the vector potential is related to the transverse electric field operator,
only the transverse electric field fluctuations would contribute to the mass
shift.

\subsection{Ratio of transverse to longitudinal electric field}

Of interest in this discussion is an estimate of how large a mass shift should
one expect if fluctuations in the transverse electric field were used instead
of fluctuations in the longitudinal electric field.
To address this, we assume for simplicity that the fluctuations scale with 
field strength (a nontrivial assumptions since fluctuations in the longitudinal
field are due to fluctuations in position, which fluctuations in the transverse
field are due to fluctuations in momentum).
If we know the ratio of the transverse to longitudinal fields near a moving charge,
then we can scale the fluctuations accordingly to develop a correction to the
mass shift estimate.

For the purposes of developing a simple scaling argument, 
the Coulomb field in the vicinity of a point charge has a magnitude of

\begin{equation}
|{\bf E}_L| ~\sim~ {e \over d^2}
\end{equation}

\noindent
where $d$ is the distance from the charge.
The magnitude of the vector potential in the vicinity of an oscillating charge is

\begin{equation}
|{\bf A}| ~\sim~ {e v \over cd}
\end{equation}

\noindent
where $v$ is the velocity of the charge.
The transverse electric field at a frequency $\omega$ is then

\begin{equation}
|{\bf E}_T| ~\sim~ {\omega e v \over c^2 d}
\end{equation}

\noindent
The ratio of the transverse field to longitudinal field is then

\begin{equation}
{|{\bf E}_T| \over |{\bf E}_L|}
~\sim~
{v \omega d \over c^2}
\end{equation}

\noindent
If the range of the moving charge is on the order of the distance with the observer

\begin{equation}
d ~\sim~ {v \over \omega}
\end{equation}

\noindent
then

\begin{equation}
{|{\bf E}_T| \over |{\bf E}_L|}
~\sim~
\left ( {v \over c} \right )^2
\end{equation}

The ratio of the mass shift estimate using the transverse electric field to that using the longitudinal electric field,
if no other feature of the model is changed, becomes

\begin{equation}
\left [ {\delta m \over m} \right ]_{CG}
~\sim~
\left [ {\delta m \over m} \right ]_{WL}
\left (
{v \over c} 
\right )^4
\end{equation}

\noindent
where the subscript CG is for Coulomb gauge, and where the subscript WL is for
Widom-Larsen.

\subsection{Oscillation frequency and scaled mass shift}

If we assume as discussed above that the relative level of fluctuations are the
same for longitudinal and transverse fields, then we need an estimate of the
proton velocity to complete the estimate.
If we adopt a high value for the oscillation frequency from neutron scattering measurements
in NbH \cite{Hauer2004}, where $\hbar \omega \sim $ 100 meV, and a large estimate
for the proton range of 1 \AA, the resulting ratio of the proton velocity to the 
speed of light is on the order of

\begin{equation}
{v \over c} ~\sim~ 5 \times 10^{-5}
\end{equation}

\noindent
In this case, the mass shift obtained using the Coulomb gauge would be on the order of

\begin{equation}
\left [ {\delta m \over m} \right ]_{CG}
~\sim~
6 \times 10^{-18}~
\left [ {\delta m \over m} \right ]_{WL}
\end{equation}

\noindent
A mass ratio of 20 estimated using a longitudinal field then would correspond to a shift in energy of less than
$10^{-10}$ eV in a Coulomb gauge calculation.

\subsection{Summary and issues}

The notion that an electron bound to a proton in a metal hydride could acquire a mass shift on the
order of an MeV due to the motion of the proton as part of collective oscillations seems highly
unlikely.
A simple way to view the effect in the Coulomb gauge can be summed up as follows.
The proton oscillates, creating a weak local magnetic field.
Fluctuations in the proton velocity then result in fluctuations in the associated magnetic field.
These fluctuations give rise to a small mass shift through Equation (\ref{shift}).

Since the local electrons can move much faster, the transverse fields developed by surface plasmon
oscillations have the potential to give rise to a larger mass shift.
Even so, such effects are tiny compared to other interactions that electrons experience in
a metal or metal hydride.

\newpage

\section{Mass shift inside a hollow current-carrying wire}

Since fluctuations in the vector potential can contribute to an
electron mass shift, we are motivated to seeks ways to increase
the effect.
In the measurements of Hollberg and Hall \cite{Hollberg1984}, the thermal shift between a
weakly bound electron (which experiences the full shift) and a more
tightly bound electron (which is shifted very little \cite{Knight1971}) was detected
as a fractional shift on the order of $2 \times 10^{-12}$ at 300 K.
At higher temperature, the mass shift is larger by the square of the temperature, 
so an increase of two orders of magnitude seems possible through heating.
However, perhaps even larger effects can be obtained through the use
of nonequilibrium conditions.
For example, it was proposed recently by Widom, Srivastava, and Larsen that
a very large mass shift could be obtained in the strong electromagnetic
fields associated with an exploding wire experiment \cite{Widom2007}.

Here, we consider a related approach in which a large current is carried in a hollow wire, 
in which a large vector potential is produced at the center.
Fluctuations in the vector potential in such a device should produce a mass shift in free (or nearly free) electrons.
This can be diagnosed spectroscopically if a gas sample is placed inside the wire.
We are interested then in maximizing fluctuations in the vector potential
in order to maximize the effect.

\subsection{Hollow wire configuration}

Although the magnetic fields associated with an exploding wire can be very large, such an experiment
may be inconvenient due to noise in the local environment, down time between shots, and the need
for a high current power source.
We seek a more subtle experimental system to work with.
 
\begin{figure} [t]
\begin{center}
\mbox{\epsfbox{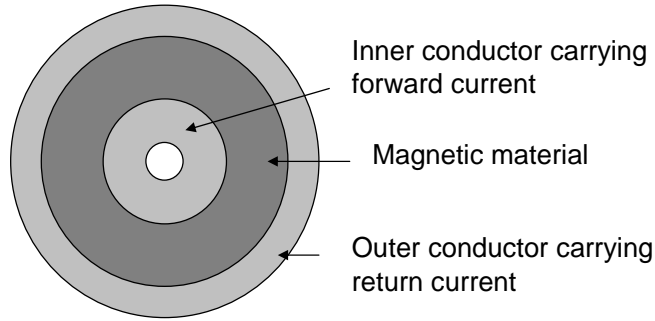}}
\caption{Hollow wire configuration.  The inner conductor (light gray) carries the forward current;
the outer conductor (light gray) carries the return current; the magnetic material in between (dark gray) maximizes
the magnetic flux density.}
\label{wire1}
\end{center}
\end{figure}

For this purpose, consider the hollow wire configuration illustrated in Figure \ref{wire1}.
The inner conductor (made up of a number of windings) carries a strong (and noisy) current 
which generates a magnetic field ${\bf H}$ within, and outside of, the conductor.
Surrounding this inner conductor is a magnetic material which serves to create a large (and noisy) magnetic flux density ${\bf B}$.
The outer conductor carries the return current, and helps confine the magnetic field.
An experimental cell can be placed inside the inner conductor for spectroscopic tests.
The vector potential inside the cell comes about as a result of the surrounding magnetic flux density.
Fluctuations in the current result in fluctuations in the magnetic field, causing fluctuations in the
magnetic flux density, producing fluctuations in the vector potential, leading ultimately to a mass shift.

\epsfxsize = 3.20in
\epsfysize = 2.00in

\subsection{The vector potential}

The magnetic field distribution in the quasi-static limit is given in Appendix A.
From the resulting magnetic flux density, the vector potential along the axis in
the center of the hollow wire configuration is calculated.
In the event that the magnetic flux density from the magnetic element of the
configuration dominates, the on-axis vector potential is

\begin{equation}
A_z(0) ~=~ {2N\mu I \over c} \ln {R_2 \over R_1}
\end{equation}

\noindent
where $N$ is the number of windings, $\mu$ is the permeability of the magnetic element, $I$ is the
current, and $R_2/R_1$ is the ratio of outer to inner radius of the magnetic material.

  This approach is conceptually simple, and is capable of generating large vector potentials.
Consider an example in which the classical current $I$ is taken to be 1 Amp (which is
$2.998 \times 10^9$ statamps since our formulas are in cgs), and the number of windings $N$ is taken to be unity.
The on-axis classical vector potential produced will be 0.1 statvolts (29.98 V) times $\mu$ (assuming the
logarithmic term is unity for this exercise).
In the case of transformer iron ($\mu$ = 4000), the resulting vector potential is 400 statvolts (120 kV).
For mu-metal ($\mu$ = 20,000), we obtain 2000 statvolts (600 kV).
We conclude that quite high vector potentials can be generated using this approach 
with only modest experimental requirements.

\subsection{Mass shift in terms of current fluctuations}

However, the mass shift is sensitive to quantum fluctuations in the vector potential, and not to
the expectation value (which corresponds to the classical estimate above).
Sizeable fluctuations are difficult to generate, as we see in the following section.
In this case, we may write 

\begin{equation}
\delta m ~=~ 
{2 N^2 e^2 \mu^2 \over  m c^6} \left ( \ln {R_2 \over R_1} \right )^2
[ \langle \hat{I}^2 \rangle - \langle \hat{I} \rangle^2 ]
\end{equation}

\noindent
which assumes a mass shift much smaller than the vacuum mass ($\delta m \ll m$).

\newpage
\section{Quantum fluctuations in a lossy driven circuit}

\epsfxsize = 3.50in
\epsfysize = 3.00in
\begin{figure} [t]
\begin{center}
\mbox{\epsfbox{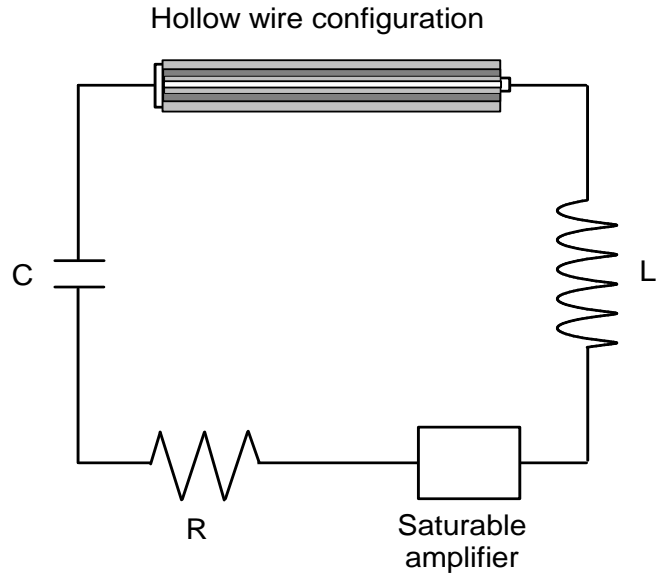}}
\caption{Circuit to supply oscillating current with large fluctuations to the hollow cylinder configuration.}
\label{circuit}
\end{center}
\end{figure}

So, under what conditions can these fluctuations be maximized?
The literature contains numerous papers concerned with the minimization of noise, but in this
case we would like to maximize the quantum noise.
One approach to the problem is to use a lossy LC-circuit with an amplifier as indicated in Figure \ref{circuit}.
This circuit is intended to implement the components that occur in a single mode laser (an oscillator, loss,
and gain that saturates), which is known to be very noisy when driven near threshold.

\subsection{Master equation}

To model intensity fluctuations in the single mode laser (and hence in this kind of circuit), 
a master equation for the photon probability distribution $p(n,t)$ has been used \cite{Mandel}
$$
{\partial \over \partial t} p(n,t)
~=~
\alpha[ np(n-1,t) - (n+1)p(n,t)]
~-~
\beta[n^2p(n-1,t) - (n+1)^2 p(n,t)]
$$
\begin{equation}
~+~
\gamma[(n+1)p(n+1,t) - n p(n,t)]
\end{equation}

\noindent
The first term on the RHS accounts for linear gain, where $\alpha$ is a gain parameter.
Gain saturation to lowest order is modeled by the second term on the RHS, where $\beta$ is a saturation
parameter.
The third term accounts for loss, where $\gamma$ is the loss parameter.

In steady state, the probability distribution satisfies

\begin{equation}
p(n) ~=~ {\alpha/\gamma \over 1 + (\beta/\alpha)n} p(n-1)
\end{equation}

\noindent
from which an exact solution can be constructed

\begin{equation}
p(n) ~\sim~ { (\alpha^2 /\beta \gamma)^n \over \Gamma[(\alpha/\beta) + n + 1] }
\end{equation}

\subsection{Fluctuations in photon number above threshold}

In \cite{Mandel} it is shown that the Stirling approximate can be used to approximate this
by a Gaussian, which can be written as

\begin{equation}
p(n) ~\sim~  e^{- {1 \over 2} (n - \langle n \rangle )^2 / \Delta n^2 }
\end{equation}

\noindent
as long as the average number of quanta $\langle n \rangle$ is much greater than the spread $\Delta n$.
The average $\langle n \rangle$ and spread $\Delta n$ in this model are given by

\begin{equation}
\langle n \rangle ~=~ {\alpha - \gamma \over \beta}
\ \ \ \ \ \ \ \ \
%
\Delta n ~=~ \sqrt{ \alpha \over \beta}
\end{equation}

Steady-state fluctuations in this model are maximized when the gain is very nearly matched by the
loss ($\alpha \approx \gamma$) for an amplifier with a very low saturation parameter $\beta$.
We can define a difference parameter $\delta$ that is the normalized difference between gain and
loss

\begin{equation}
\delta ~=~ {\alpha - \gamma \over \alpha}
\end{equation}

\noindent
In terms of this parameter, the relative fluctuations are

\begin{equation}
{ \Delta n \over \langle n \rangle }
~=~
{1 \over \sqrt{ \langle n \rangle  \delta}}
\end{equation}

\noindent
The spread $\Delta n$ for a classical state is simply $\sqrt{\langle n \rangle}$.
The fluctuations here are larger by a factor of $1/\sqrt{\delta}$ due to the 
diffusion in $n$ associated with the loss and gain in the master equation.

%
%
%


\subsection{Current fluctuations above threshold}

\epsfxsize = 3.80in
\epsfysize = 2.80in
\begin{figure} [t]
\begin{center}
\mbox{\epsfbox{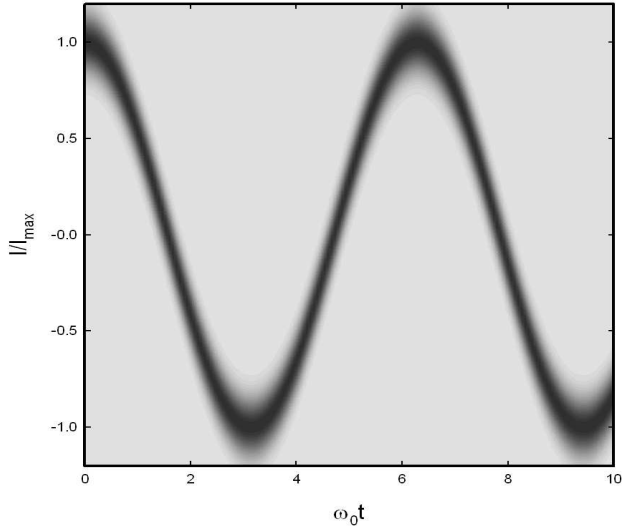}}
\caption{Sinusoidal signal with fluctuations.}
\label{cosine}
\end{center}
\end{figure}

Fluctuations in the current can be determined from fluctuations in number and phase through
a perturbative approach.
The expectation value of the current can be related to expectation values of number $\hat{n}$ and phase $\hat{\phi}$
according to

\begin{equation}
\langle \hat{I} \rangle 
~=~
\sqrt{\hbar \omega_0 \langle \hat{n} \rangle \over 2 L} \cos ( \omega_0 t + \langle \hat{\phi} \rangle)
\end{equation}

\noindent
We use $L$ for the total inductance of the circuit, and where the characteristic frequency of the
circuit $\omega_0$ satisfies

\begin{equation}
\omega_0^2 ~=~ {1 \over LC}
\end{equation}

\noindent
where $C$ is the capacitance.
If the fluctuations are small relative to the average (as depicted in Figure \ref{cosine}), then
we can linearize around the sinusoid to obtain

\begin{samepage}
{\small

$$
\hat{I} - \langle \hat{I} \rangle 
~=~
{1 \over 2}
\sqrt{\hbar \omega_0 \langle \hat{n} \rangle \over 2 L} \cos ( \omega_0 t + \langle \hat{\phi} \rangle) 
\left ({\hat{n} - \langle \hat{n} \rangle \over \langle \hat{n} \rangle} \right )
\ \ \ \ \ \ \ \ \ \ \ \ \ \ \ \ \ \ \
\ \ \ \ \ \ \ \ \ \ \ \ \ \ \ \ \ \ \
$$
\begin{equation}
\ \ \ \ \ \ \ \ \ \ \ \ \ \ \ \ \ \ \
-
\sqrt{\hbar \omega_0 \langle \hat{n} \rangle \over 2 L} \sin ( \omega_0 t + \langle \hat{\phi} \rangle)
(\hat{\phi} - \langle \hat{\phi} \rangle)
\end{equation}

\noindent
The fluctuations in current are then given by

$$
\langle (\hat{I} - \langle \hat{I} \rangle)^2 \rangle
~=~
{1 \over 4}
\left ( 
{\hbar \omega_0 \langle \hat{n} \rangle \over 2 L} 
\right )
\cos^2 ( \omega_0 t + \langle \hat{\phi} \rangle) 
\left [ 
{ \langle (\hat{n} - \langle \hat{n} \rangle)^2 \rangle \over \langle \hat{n} \rangle^2} \right ]
\ \ \ \ \ \ \ \ \ \ \ \ \ \ \ \ \ \ \ \ \ \ \ \ \ \ \ \ \
$$
$$
-
\left ({\hbar \omega_0 \langle \hat{n} \rangle \over 2 L} \right )
\cos ( \omega_0 t + \langle \hat{\phi} \rangle) 
\sin ( \omega_0 t + \langle \hat{\phi} \rangle)
\left [ {(\hat{n} - \langle \hat{n} \rangle)(\hat{\phi} - \langle \hat{\phi} \rangle)  \over \langle \hat{n} \rangle} \right ]
$$
\begin{equation}
+
\left ( {\hbar \omega_0 \langle \hat{n} \rangle \over 2 L} \right )
 \sin^2 ( \omega_0 t + \langle \hat{\phi} \rangle)
\langle (\hat{\phi} - \langle \hat{\phi} \rangle)^2 \rangle
\end{equation} 

}
\end{samepage}

\noindent
The current fluctuations are due to number fluctuations alone when

$$ \sin ( \omega_0 t + \langle \hat{\phi} \rangle) ~=~ 0$$

\noindent
In this case we obtain

\begin{equation}
\langle (\hat{I} - \langle \hat{I} \rangle)^2 \rangle
~=~
{1 \over 4}
\langle \hat{I} \rangle_{max}^2
\left [ 
{ \langle (\hat{n} - \langle \hat{n} \rangle)^2 \rangle \over \langle \hat{n} \rangle^2} \right ]
\end{equation}

\subsection{Current fluctuations below threshold}

When run below threshold, the excitation of the oscillator is much weaker, so that we can
neglect saturation.
In this case we obtain a thermal distribution in steady state

\begin{equation}
p(n) 
~=~ 
{\alpha \over \gamma} ~p(n-1)
~=~
e^{- \hbar \omega_0/k_B T_{eff}}~p(n-1)
\end{equation}

\noindent
where $T_{eff}$ is the effective temperature associated with the amplifier and loss.
In the high temperature limit, the current fluctuations are

\begin{equation}
\langle \hat{I}^2 \rangle - \langle \hat{I} \rangle^2 
~=~
{\hbar \omega_0 \over L} \left [ {1 \over e^{\hbar \omega_0/k_B T_{eff}} - 1} + {1 \over 2} \right ]
~\to~
{k_B T_{eff} \over L} 
\end{equation}

\subsection{Mass shift estimates}

We can use these results to develop estimates for the mass shift.  
Above threshold, the current fluctuations appear in connection with an oscillating
signal, and are limited by how closely the gain matches the loss.
Below threshold, only fluctuations occur, and the effective noise temperature is
determined once again by how closely the gain matches the loss.
In the circuit that we examined here, the gain and loss are variable, but in the
steady-state solutions we assumed that they remain fixed.
In practice, one would use a more sophisticated arrangement with a very hot resistive
element to inject noise (not included in our master equation), and feedback to 
keep the gain and loss closely matched (not included in our model).
Consequently, it makes sense here to characterize the noise in terms of an 
effective temperature in order to evaluate the magnitude of the mass shift and
associated energy shift.

From this discussion, we may write the mass shift in terms of the effective temperature
for the below-threshold case as

\begin{equation}
\delta m ~=~ 
{2 N^2 e^2 \mu^2 k_B T_{eff} \over  m c^6 L} \left ( \ln  {R_2 \over R_1}  \right )^2
\ \ \ \ \ \ \
({\rm below~threshold})
\end{equation}

\noindent
Before continuing, we note that mass shift is maximized when the inductance is minimized,
so that no additional inductance should be used.
In this case, the total inductance for the circuit is very nearly that of the hollow
wire ($L_{hw}$), which is given in Appendix A to be

\begin{equation}
L_{hw} ~=~ {2 N^2 \mu l_z \over c^2} \ln {R_2 \over R_1}
\end{equation}

\noindent
where $l_z$ is the length of the hollow wire.
Upon inserting, we obtain for the mass shift

\begin{equation}
\delta m ~=~ 
{e^2 \mu k_B T_{eff}  \over  m c^4 l_z} \ln  {R_2 \over R_1} 
\ \ \ \ \ \ \
({\rm below~threshold})
\end{equation}

\noindent
The associated energy shift evaluates to

\begin{equation}
\delta m c^2
~=~
5.63 \times 10^{-10} ~{\rm eV}~
\ln  {R_2 \over R_1}
\left [{10~{\rm cm} \over l_z} \right ]
\left [ {\mu \over 20000} \right ]
\left [ { k_B T_{eff} \over 1~{\rm eV}} \right ]
\end{equation}

\noindent
Based on this, we would expect that energy shifts in the range of $10^{-9}$ to
$10^{-6}$ eV should be possible by maximizing the noise temperature, and by
taking advantage of more advanced magnetic materials.

\newpage

\section{Summary and Conclusions}

  Our effort was stimulated by the recent publication of Widom and Larsen who proposed that
a large electron mass shift could be expected to occur near the surface of a metal hydride \cite{Widom2006}.
These authors were led to this conclusion from a gauge-free formulation of the mass shift.
Since the result is so counter to our intuition, we decided to investigate making use
of an approach based on the Coulomb gauge.
In the end, the key difference is that in the Coulomb gauge one must use the transverse
electric field fluctuations for a mass shift estimate instead of fluctuations in the
longitudinal field.
As a result, the mass shift that we would expect would be orders of magnitude smaller.

As a result of the Widom and Larsen proposal, we were motivated to consider the problem
of creating a more significant mass shift than can be obtained thermally, by using
nonequilibrium conditions to maximize the fluctuations in the potential vector.
To this end, we proposed the use of a hollow wire with a magnetic element driven by a noisy current source;
the mass shift in such a configuration is proportional to the current fluctuations of
the circuit.
To maximize current fluctuations, we considered a lossy LC circuit with an amplifier,
which is closely related to the problem of a single mode laser which is known to
be extremely noisy near threshold.
Fluctuations in the oscillator photon number were modeled using a simple master equation borrowed
from laser physics, which has been used to describe the photon distribution in a single mode laser.
Above threshold the signal is oscillatory, with small fluctuations which are maximized
near threshold.
Below threshold, the photon distribution is thermal at a temperature determined by
the ratio of the gain to the loss;
near threshold, the gain very nearly matches the loss, and the effective temperature
can be very high.

To maximize the mass shift, the circuit inductance should be minimized.
When the inductance of the hollow wire dominates the circuit inductance,
the below-threshold mass shift depends on the permeability, the geometry of the cylinder, and on
the noise temperature.
It is independent of the number of windings.
The mass shift induced in this way is sufficiently large to be observable (we believe that
energy shifts in the range of $10^{-9}$ to $10^{-6}$ eV can be produced in a free, or weakly-bound,
 electron), and it can be greater than the thermal shift observed previously
near room temperature.
The development of very high noise temperatures in the circuit (in the keV range or higher) will require
the use of a more sophisticated circuit than the one analyzed here, since it is difficult to
match gain and loss so precisely without feedback.

Note that the mass shift produced in such an experiment occurs under conditions where the
classical electric and magnetic fields are zero [in the above threshold case, we are focused
on the $\sin(\omega t+\langle \hat{\phi} \rangle)=0$ condition in which the expectation
value of the transverse electric field is zero].
It appears as if the electron is exhibiting a response to the vector potential in such an
experiment.

\newpage
\appendix
\section{Classical vector potential estimate}

\begin{figure} [t]
\begin{center}
\mbox{\epsfbox{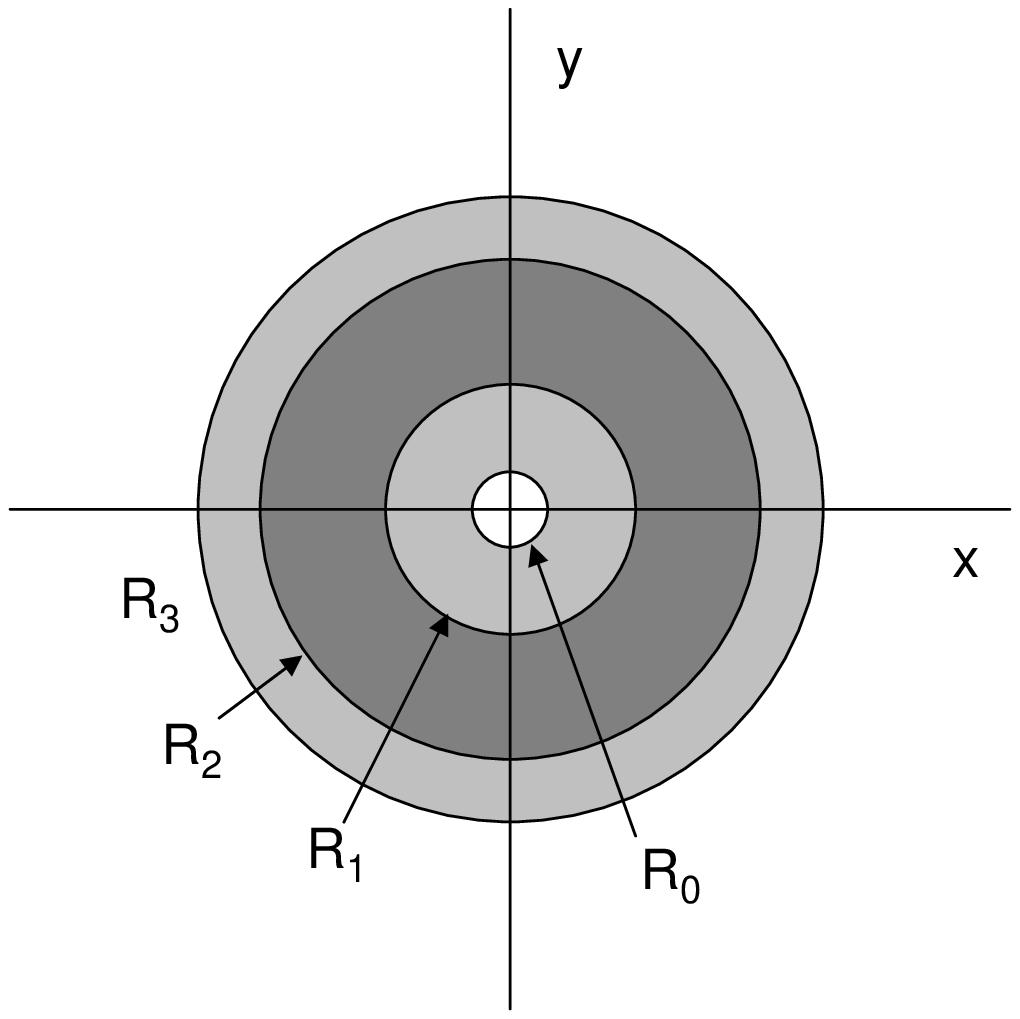}}
\caption{Cross section of a simple hollow wire configuration.  The center is hollow, out to a radius $R_0$ (indicated in white).
An inner conductor is shown between $R_0$ and $R_1$ (in light gray).
A magnetic material with permeability $\mu$ is indicated between $R_1$ and $R_2$ (in dark gray). 
An outer nonmagnetic conductor to carry the return current is illustrated between $R_2$ and $R_3$ (in light gray).}
\label{wire}
\end{center}
\end{figure}

In this appendix we consider the vector potential due to a simple hollow wire configuration
as illustrated in Figure \ref{wire}.
The innermost hollow cylinder (between $R_0$ and $R_1$) is a conductor for the forward current, assumed to be made
from a nonmagnetic metal such as copper.
This conductor is surrounded by a magnetic material (between $R_1$ and $R_2$) such as iron or permalloy.
An outer cylindrical conductor is present to carry a return current.

\epsfxsize = 3.20in
\epsfysize = 2.80in

\subsection{Magnetic field}

If we assume that the current is carried uniformly in the nonmagnetic conductors, and that the system is
magneto-quasistatic, then we can estimate the magnetic field using

\begin{equation}
\oint_C {\bf H} \cdot d{\bf l}
~=~
{4 \pi \over c} \int {\bf J} \cdot \hat{\bf n} ~d^2a
\label{ampere}
\end{equation}

\noindent
using circular contours at different radial distances $\rho$ away from the center.
We assume that the current in the inner hollow cylinder is $z$-directed 

\begin{equation}
{\bf J} ~=~ \hat{\bf i}_z J_0 \ \ \ \ \ \ ({\rm inner~conductor})
\end{equation}

\noindent
The outer cylinder carries the return current, and is also $z$-directed

\begin{equation}
{\bf J} ~=~ -\hat{\bf i}_z J_1 \ \ \ \ \ \ ({\rm outer~conductor})
\end{equation}

\noindent
The magnetic field is $\phi$-directed, and can be estimated simply from Equation (\ref{ampere})

\begin{equation}
{\bf H}(\rho)
~=~
\left \lbrace
\begin{array} {ll}
0   & 0 \le \rho \le R_0 \cr
\hat{\bf i}_\phi \displaystyle{2 \pi J_0 \over \rho c} (\rho^2 - R_0^2) & R_0 \le \rho \le R_1 \cr
\hat{\bf i}_\phi \displaystyle{2 \pi J_0 \over \rho c} (R_1^2  - R_0^2) & R_1 \le \rho \le R_2 \cr
\hat{\bf i}_\phi \left [
\displaystyle{2 \pi J_0 \over \rho c} (R_1^2  - R_0^2)-\displaystyle{2 \pi J_1 \over \rho c} (\rho^2  - R_2^2) 
\right ]
& R_2 \le \rho \le R_3 \cr
0 & \rho > R_3 \cr
\end{array}
\right .
\end{equation}

\noindent
The magnetic field is zero outside since the return current matches the forward current

\begin{equation}
J_0 \pi (R_1^2-R_0^2) ~=~ J_1 \pi (R_3^2 - R_2^2)
\end{equation}

\subsection{Vector potential}

The vector potential at the $z$-axis can be found from

\begin{equation}
\oint_{C'} {\bf A}\cdot d{\bf l} ~=~ \int \mu {\bf H} \cdot d^2a
\end{equation}

\noindent
For this we use a rectangular contour that travels a short distance $\Delta z$ along the $z$-axis; then 
radially outward beyond the outer conductor; then backward the same distance in $z$; and then
radially inward.
Since the magnetic field is $\phi$-directed, no contribution is obtained for the radial legs.
Hence, we obtain

$$
[A_z(0)-A_z(R_3)]\Delta z 
~=~
\Delta z 
\int_{R_0}^{R_1} {2 \pi J_0 \over \rho c} (\rho^2 - R_0^2) d \rho
+
\Delta z
\int_{R_1}^{R_2} {2 \pi \mu J_0 \over \rho c} (R_1^2  - R_0^2) d \rho
$$
\begin{equation}
+
\Delta z
\int_{R_2}^{R_3} 
{2 \pi J_0 \over \rho c} (R_1^2  - R_0^2)
-
{2 \pi J_1 \over \rho c} (\rho^2  - R_2^2) 
d \rho
\end{equation}

\noindent
Integrating results in

{\small

\begin{equation}
A_z(0) 
~=~
\left ( {2I \over c} \right )
\left [
\mu \ln {R_2 \over R_1}
+ 
{1 \over 2}
+
\ln {R_3 \over R_2}
- 
{R_0^2 \over R_1^2-R_0^2} \ln {R_1 \over R_0}
+
{R_2^2 \over R_3^2-R_2^2} \ln {R_3 \over R_2}
\right ]
\end{equation}

}

\noindent
where we have assumed that the vector potential $A_z(R_3)$ is zero outside the outermost conductor.

In the event that the magnetic permeability $\mu$ of the magnetic section is much greater
than unity, then the contribution of the magnetic material dominates.
In this case, we may write

\begin{equation}
A_z(0) ~=~ {2\mu I \over c} \ln {R_2 \over R_1} 
\end{equation}

\subsection{Looped wire}

\begin{figure} [t]
\begin{center}
\mbox{\epsfbox{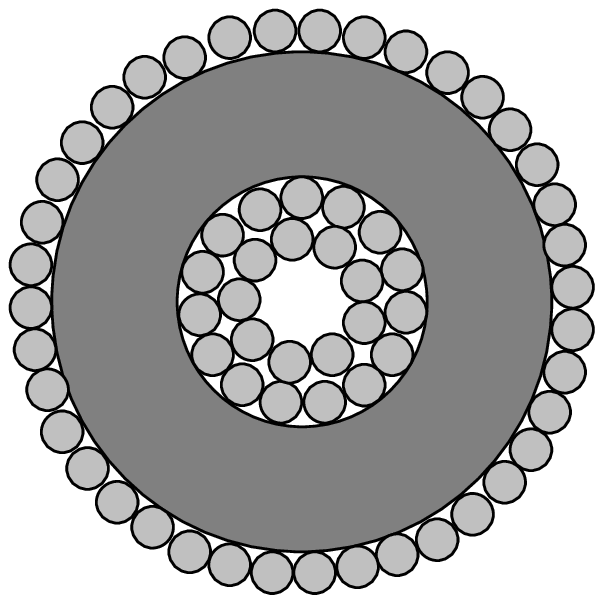}}
\caption{Cross section a looped configuration.  A wire (light gray) is looped around a cylindrical shell of
magnetic material (dark gray) $N$ times.  A single wire carries the current $I$, but the total current
passing inside the magnetic material is $NI$.}
\label{loop}
\end{center}
\end{figure}

In the event that a high-$\mu$ material is used, then the ratio of vector potential to current
can be increased by looping a wire carrying the drive current around the magnetic element,
as indicated in Figure \ref{loop}.  The vector potential on axis in this case is

\begin{equation}
A_z(0) ~=~  {2 N \mu I \over c} \ln {R_2 \over R_1}
\end{equation}

\noindent
where $N$ is the number of windings.
The self-inductance of the hollow wire $L_{hw}$ is

\begin{equation}
L_{hw} ~=~ {2 N^2 \mu l_z \over c^2} \ln {R_2 \over R_1}
\end{equation}

\noindent
where $l_z$ is the length of the hollow wire.

\epsfxsize = 3.20in
\epsfysize = 2.80in

\end{document}